# Error analysis of vertical test for CEPC 650 MHz superconducting radio-frequency cavity

Lingxi Ye[a,b,c], Peng Sha [a,b,c]*, Zhenghui Mi [a,b,c]**, Feisi He [a,b,c], and Jiyuan Zhai [a,b,c]

[a] *Institution of high energy physics, Chinese Academy of Sciences, Beijing 100049, China*

[b] *School of Nuclear Sciences and Technology, University of Chinese Academy of Sciences, Beijing 100049, China*

[c] *Center for Superconducting RF and Cryogenics, Institute of High Energy Physics, Chinese Academy of Sciences, Beijing 100049, China*

*e-mail: shapeng@ihep.ac.cn

**e-mail: mizh@ihep.ac.cn

**Abstract**—Hundreds of 650 MHz superconducting radio-frequency (SRF) cavities with high intrinsic quality factor ($Q_0$) and accelerating gradient ($E_{acc}$) will be adopted for Circular Electron Positron Collider (CEPC). The values of $Q_0$ and $E_{acc}$ are obtained during vertical test at 2.0 K. Hence, high accuracy of vertical test is essential for evaluating the performance of SRF cavity. The 650 MHz SRF cavities achieved very high $Q_0$ ($6 \times 10^{10}$) and $E_{acc}$ (40 MV/m) during the vertical test. In our study, the error analysis of vertical test was conducted in the scalar case, in order to achieve high accuracy. The uncertainties of vertical test were obtained through calculation, which was approximately 3% for $E_{acc}$ and less than 5% for $Q_0$. This result was reasonable and acceptable.

## INTRODUCTION

In the last decades, SRF cavities with high $E_{acc}$ and $Q_0$ have been adopted extensively by large accelerators, such as high-energy colliders, free electron lasers, synchrotron photon sources and neutron sources[1-7]. The measurements of $Q_0$ and $E_{acc}$ are implemented through the vertical test, during which SRF cavities are immersed in liquid helium of 2.0 or 4.2 K. The values of $Q_0$ and $E_{acc}$ are originated from the calculation of RF power, which is incident into the cavity, reflected and extracted from the cavity. Due to advanced recipes of surface process adopted (electro-polishing, nitrogen doping/infusion, medium-temperature baking, et al), $Q_0$ and $E_{acc}$ of SRF cavities has been greatly improved recently [8–12]. Hence, it's very important to ensure the accuracy of the vertical test, which will help enhancing the performance and understanding of SRF cavities.

There are many vertical test stands (VTS) for SRF cavities worldwide [13-15], where various SRF cavities receive tests. As for the SRF cavities with high $Q$ and high gradient, the higher accuracy of vertical test should be achieved. Therefore, in recent years, the study about the error and uncertainty of vertical test has been widely carried out. The theory of vertical test of SRF cavities is introduced, and preliminary error calculation is also demonstrated at Thomas Jefferson National Accelerator Facility (Jefferson Lab) in Ref.[13]. The calculation taking into account the correlation between power meters is introduced in Ref.[16]. And the additional error calculations for the upgraded VTS at Jefferson Lab are also mentioned, which indicate the influence of the coupling during the τ-measurement on the final error[17]. Besides, the related research has been conducted at many other SRF laboratories[18-23].

This paper focuses on the vertical test of the 650 MHz SRF cavity at the Institute of High Energy Physics, Chinese Academy of Sciences (IHEP), where the error is analyzed in the scalar case. The reminder is organized as follows. First introduces the VTS at IHEP, which includes test theory, loop composition, test procedure, and calibration. Then, the system uncertainty of VTS and error calculations on the test results of the 650 MHz SRF cavity are investigated. Finally, a brief conclusion is presented.

## SETUP AND PROCESS OF VETTICAL TEST

Due to the ultra-high $Q_0$ @ 2.0 K, the bandwidth of the 650 MHz SRF cavity is even less than one Hz during the vertical test. Meanwhile, the frequency of SRF cavity is affected by various factors, such as mechanical vibration, Lorentz detuning, et al. Therefore, it's important to accomplish fast tracking of the frequency and phase during the vertical test of 650 MHz SRF cavity. There are two main technical schemes: one is voltage controlled oscillators configured in phase locked loops (VCO-PLL) [13], the other is self-excited loop (SEL) with positive feedback[22].

Compared with VCO-PLL, SEL has simpler structure and can achieve faster resonance, which is regarded as a reasonable option and adopted by the VTS at IHEP. SEL mainly consists of a high-power RF amplifier, a phase shifter and a limiter. When the gain of the loop is > 0 dB, the signal is self-excited and amplified until be limited by the limiter. Equation (1) shows the overall phase shift of the loop, where $\varphi_c$ denotes phase shift of SRF cavity, $\varphi_s$ denotes phase shift of phase shifter, $\varphi_{path}$ denotes the phase shift in the rest of the loop. $\omega_c$ denotes the cavity resonant frequency and $\omega$ denotes the excitation signal frequency. By adjusting the phase of the phase shifter manually, $\varphi_c$ could be changed. When the loop in the excitation frequency coincides with the SRF cavity excitation frequency, the SRF cavity is on resonance and the RF signal probed is maximized. Hence, the oscillation frequency of the whole loop is actually determined by the resonant frequencies of SRF cavities.

$$\varphi_c(\omega_c, \omega) + \varphi_{path}(\omega) + \varphi_s(\omega) = 2n\pi \quad (1)$$

The schematic and control cabinets of VTS at IHEP are shown as Fig.1 and Fig.2, respectively. In addition to the key components of SEL, there are also two directional couplers, one circulator, three RF-power dividers, two dual-channel power meters, one frequency meter, one spectrometer, et al. All these components are connected to each other by high-quality coaxial cables. Specially, instead of just one directional coupler, combination of two directional couplers and one circulator is used to extract the $P_{in}$ and $P_{rm}$ signals at the input port. It effectively blocks signals reflected from the cavity to the incident port of the power meter, which may result to a large measurement error as mentioned in Ref.[19], At the same time, this design avoids measurement errors caused by insufficient isolation, which occurs using just one directional coupler.

The process of vertical test consists of three steps: cable calibration, decay time measurement and continuous-wave (CW) measurement. During cable calibration, the attenuation coefficients $C_{in}$, $C_r$, and $C_t$ from the power meter to the cavity are calibrated, aiming to provide an accurate power reading at the cavity entrance with $P_x = C_x P_{xm}$. Here $P_x$ means the power at the cavity port, $C_x$ is the corresponding calibration coefficient, and $P_{xm}$ is the measured value of the power meter. During the vertical test, the SRF cavities are commonly immersed with liquid helium in the cryogenic dewar. Hence, a section of the RF cable is in the cryogenic dewar, which could only be calibrated by single-end. The RF cables need to be disconnected at point B,C,D for separate attenuation calibrations as Fig.1. In addition, the cable from the control room to the cryogenic dewar is too long to be measured directly using the network analyzer. To measure $C_t$, the Pickup cable is disconnected at point C and D. The return loss of the cable in the dewar is measured through the network analyzer. An additional signal source and power meter are used to measure the attenuation from the dewar to the control room through $P_t$ and $C_t$, which is defind as Equation (2). Here, $P_{dewar}$ denotes the power monitored by the additional power meter at the top of the dewar with an external signal source, which outputs with fixed power (typically 12 dBm). $P_{tm}$ denotes the value of the $P_t$ in the control room. Since the value of $S_{11}$ measured through the network analyzer indicates the reflect loss of the cable in the dewar, the transmission loss should be divided by 2.

$$C_t = 10^{-(P_{tm} - P{\text{dewar}} - 1/2|S_{11}|)/10}. \quad (2)$$

$C_{in}$ and $C_r$ are obtained at the same time. As shown in Fig.1, the transmission cable is firstly disconnected at point A, which is between the circulator and dewar. The amplifier's excitation is adjusted so that the power measured at point A is around 80 mW (close to upper limit of the power meter), denoted as $P_d$. At the same time, the value of the input power meter is recorded as $P_0$. Then, we connect a short termination to port A, change the amplifier's excitation power until the input power meter reaches $P_0$. At this condition, the reading on the reflected power meter is recorded as $P_{rd}$. Finally, point A is connected to dewar through cable, restoring the complete wiring setup. After adjusting the excitation signal to make the value of input power meter be $P_0$ again, the reading on the reflected power meter is denoted as $P_{rcable}$. If there are multiple pick up ports, only additional calibration for $C_t$ is required for the extra ports, more calibration calculations for $C_{in}$ and $C_r$ are unnecessary, which are denoted as Equation (3-4) respectively.

$$C_{in} = \frac{P_d}{P_0}\sqrt{\frac{P_{rcable}}{P_{rd}}}. \quad (3)$$

$$C_r = \frac{P_d}{P_{rd}}\sqrt{\frac{P_{rd}}{P_{rcable}}}. \quad (4)$$

The decay time measurement aims to calculate the external quality factor ($Q_t$) of the pickup antenna under critical coupling. The decay time is obtained using the half-power method, which measure the time for cavity power to drop by 3 dB through a signal analyzer, as Equation (5). By measuring the decay time constant, the corresponding $Q_L$ and the values of $Q_t$ and $Q_0$ could be calculated through Equation (6-8).

$$\tau = \frac{\tau_{1/2}}{\ln 2} \quad (5)$$

$$Q_L = 2\pi f_0 \tau \quad (6)$$

$$Q_t = 4\pi f_0 \tau \frac{P_{in} + C\sqrt{P_{in}P_r}}{P_t} \quad (7)$$

$$Q_0 = (1 + \beta_1 + \beta_2) \cdot Q_L, \quad (8)$$

Here, $C$ indicates the coupling state of the cavity. $C=1$ means over coupling, while $C=-1$ under coupling. $\beta_1$ and $\beta_2$ represent the coupling coefficients of the input and pickup antennas,
$$\beta^* = (\sqrt{P_{in}} + C\sqrt{P_r}) / (\sqrt{P_{in}} - C\sqrt{P_r}).$$
respectively.and they are related by $\beta_1=\beta^*(1+\beta_2)$, where

For most vertical test scenarios, the coupling coefficient of the pickup antenna is 1-2 orders of magnitude lower than that of the input antenna. Therefore, when calculating $Q_0$, the contribution of the pickup antenna's coupling coefficient can be ignored. The value of $Q_t$ obtained under critical coupling, which remains constant at different RF fields, can be used for $Q_0$ calculation during the CW measurement. By measuring the corresponding RF power ($P_t$, $P_{rm}$ and $P_{in}$), $E_{acc}$ and $Q_0$ can be determined.

$$E_{acc} = \sqrt{Q_t P_t \frac{R/Q}{L}} \quad (9)$$

$$Q_0 = Q_t P_t / P_{loss}, \quad (10)$$

Here, $P_{loss}=P_{in}-P_t-P_r$ represents the RF power dissipated in the cavity. For the 650 MHz SRF cavity studied in this paper, $R/Q$ is 105 Ω and $L$ is 0.23 m, which is half of wave length.

### ERROR ANALYSIS OF 650MHz SRF CAVITY

The analysis of errors should base on specific testing equipment and testing processes. For the VTS mentioned above, we perform error analysis in two steps. Firstly, statistical methods were used to obtain the error during measurements directly reading from the instruments. Secondly, the error during indirect measurements were calculated by using error propagation.

For the uncertainties in direct measurements, analysis should be conducted with considering both Type A uncertainties and Type B uncertainties. The uncertainties primarily stemmed from the RF power monitored by the power meter ($P_m$), the decay time constant ($\tau$), and the resonant frequency ($f$). The type of power meters measuring $P_{im}$, $P_{rm}$, and $P_{tm}$ was KEYSIGHT E4415A EPM-P series, which adopted the average mode and was equipped with E9321A type power probes. In addition, once the data of RF power was transferred to the computer, it was averaged again. The corresponding Type A uncertainty was 0.5% after calculations. The power meter was operated at ~ 25 ℃, its non-linear error was 4%. The measurement accuracy was 0.5%, and the uncertainty of the power probe's calibration coefficient was 1%. Thus, the total Type B uncertainty was 4.2%. Using the combined uncertainty equation, the direct measurement uncertainty for power was determined: 4.2%.

Similarly, the value of $\tau$ was measured directly through the Tektronix RSA5106A real-time signal spectrum analyzer, whose Type B uncertainty was 4%[20]. The uncertainty induced by the $Q$-slope was disregarded in this study.

The uncertainties of RF power related to the power meters pertained to their accuracy during measurement of incident power. However, due to mismatches in the microwave circuitry, the RF power monitored by these power meters might not be entirely accurate. For instance, if the directivity of circulator was suboptimal, part of the RF power reflected from the SRF cavity might reach the input port. And the incident power could reach the reflected power detection port. To mitigate it, the combination of double directional couplers and circulators in the microwave circuitry was adopted. With a full reflection termination connected to port 2 of the circulator, the insertion loss at port 2 of the reflected directional coupler and port 2 of the incident directional coupler was -66 dB. It indicated that such a combination of double directional couplers and circulators reduced crosstalk between forward and reflected RF effectively.

Except the direct measurements mentioned above, all the other parameters were obtained indirectly. As listed in Equation(11), the uncertainty of the indirect measurement, $U_f(x_1...x_n)$ was evaluated based on the uncertainties of the error sources ($U_{xi}$), and their correlation coefficient $r(x_i, x_j)$. Hence, it's essential to consider the correlation of error sources.

$$U_f(x_1..x_n) = \sqrt{\sum_{i=1}^{n}(\frac{\partial f}{\partial x_i} \cdot U_{x_i})^2 + 2\sum_{i=1}^{n-1}\sum_{j=i+1}^{n}\frac{\partial f}{\partial x_i}\frac{\partial f}{\partial x_j}r(x_i, x_j)u(x_i)u(x_j)}. \quad (11)$$

The correlation coefficients were according to two principles. Firstly, data from the same instrument was linearly correlated. For instance, the input power $P_{in}$ measured during the measurement of $\tau$ and the CW measurement were linearly related, namely correlation coefficient = 1. Secondly, For two values monitored simultaneously, their correlation coefficient was demonstrated as Equation(12). It was assumed that the remaining values were linearly independent.

$$r(x_i, y_i) = \frac{\sum_{i=1}^{n}(x_i - \bar{x})(y_i - \bar{y})}{(n-1)s(x)s(y)}. \quad (12)$$

Equation (13-14) were used to facilitate the calculation of correlation coefficients. Here, $P'_{im}$, $P'_{rm}$, and $P'_{tm}$ denoted the values from power meter during the CW measurement and $P_{im}$, $P_{rm}$, $P_{tm}$ the values during the measurement of $\tau$. The variables influencing $Q_0$ and $E_{acc}$ are : $f$, $\tau$, $C$, $C_{in}$, $C_r$, $C_t$, $P_{im}$, $P_{rm}$, $P_{tm}$, $P'_{im}$, $P'_{rm}$, $P'_{tm}$, $R/Q$ and $L$. Although the values of $C_{in}$, $C_r$ and $C_t$ are not directly obtained by the instrument, their determination is only during cable calibration, does not change in other testing process, and they are not related to other variables. Values of $R/Q$ and $L$ are determined during design and are considered unchanged during testing. Therefore, they are considered independent input quantities. By constructing a covariance matrix $M$(15) of these variables, error calculations were carried out conveniently. The diagonal elements of this matrix represented the variance of each variable, while the off-diagonal elements represented the covariance between variables. Finally, the uncertainties of $Q_0$ and $E_{acc}$ were calculated using Equation (16). $\nabla f$ denoted the gradient matrix of $Q_0$ and $E_{acc}$ observed with respect to all input variables as $\nabla f = [\frac{\partial f}{\partial x_1}, \frac{\partial f}{\partial x_2}, \ldots, \frac{\partial f}{\partial x_n}]$, while $\nabla f^T$ was its transposed matrix.

$$Q_0 = 4\pi f \tau \frac{C_{in}P_{im} + C_\beta \sqrt{C_{in}P_{im}C_r P_{rm}}}{C_t P_{tm}} \frac{C_t P'_{tm}}{C_{in}P'_{im} - C_t P'_{tm} - C_r P'_{rm}} \quad (13)$$

$$E_{acc} = \sqrt{4\pi f \tau \frac{C_{in}P_{im} + C_\beta \sqrt{C_{in}P_{im}C_r P_{rm}}}{C_t P_{tm}} C_t P'_{tm} \frac{r/Q}{L}} \quad (14)$$

$$M = \begin{bmatrix} u_{x_1}^2 & ru_{x_1}u_{x_2} & \cdots & ru_{x_1}u_{x_n} \\ ru_{x_2}u_{x_1} & u_{x_2}^2 & \cdots & ru_{x_2}u_{x_n} \\ \vdots & \vdots & \ddots & \vdots \\ ru_{x_n}u_{x_1} & ru_{x_n}u_{x_2} & \cdots & u_{xn}^2 \end{bmatrix}. \quad (15)$$

$$(\delta f)^2 = \nabla f^T \cdot M \cdot \nabla f \quad (16)$$

The values of variables used are listed in Table 1, which were adopted for the error analysis of the 650 MHz SRF cavity. The decay time was 3.3 s at $P_{in}$=0.13 watt, $Q_t$ was $5.4 \times 10^{11}$, $Q_{in}$ was $2.6 \times 10^{10}$ and the VSWR was 2.266 at 2.0 K. The vertical test results at 2.0 K and 1.9 K with error bar are shown as Fig.3, which adopted the same parameters of cable calibration and decay time measurement. The uncertainty of $Q_0$ at 1.9 K was greater than 2.0 K, because of further deviation from critical coupling. The errors of $C_{in}$, $C_r$, $C_t$ were obtained from the corresponding measurement during the cable calibration, according to the error synthesis formula. The uncertainty evaluation of vertical test results (at 2.0 K) is shown as Fig.4. The uncertainty of $E_{acc}$ predominantly depended on $\Delta Q_t$ and $\Delta P'_{tm}$, both of which were unchanged during the CW measurement. So, the fractional $E_{acc}$ uncertainty almost remained constant (approximately 3%) as a function of $E_{acc}$. Meanwhile, the uncertainty of $Q_0$ trends in line with $\beta^*$ as a function of RF field. The reason is discussed here briefly. As the accelerating gradient increased, $Q_0$ decreased gradually. It resulted in reduction of $\beta^*$, which slided from over coupling to critical coupling. It led to gradual decrease of the fractional $Q_0$ uncertainty, which was always less than 5%. At higher RF fields, $Q_0$ continued to decrease and $\beta^*$ slided from critical coupling to under coupling. Hence, the uncertainty of $Q_0$ and $\beta^*$ turned around and began to increase.

## CONCLUSIONS

In this paper, setup and process of vertical test for CEPC 650 MHz SRF cavity was introduced. Scalar error was investigated through analyzing the uncertainty of direct and indirect measurement values, taking into account the correlation between input quantities. Even so, the small difference of couplings could result in the big difference of the uncertainties. Hence, it was required to predict the coupling of the input antenna. Although the measurement accuracy of microwave circuits was improved through loop optimization, it had little effect. Next, the VTS at IHEP would be upgraded. The advantages of vector test and digital acquisition will be utilized. Therefore, higher accuracy could be achieved under various couplings.

## FUNDING

This work was supported by the National Natural Science Foundation of China (No. 12075270), the Strategic Priority Research Program of the Chinese Academy of Sciences (No. XDB25000000), and the Platform of Advanced Photon Source Technology R&D.

## REFERENCES

1. H. Padamsee, 50 years of success for srf accelerators—a review, Superconductor Science and Technology 30 (2017) 053003. https://dx.doi.org/10.1088/1361-6668/aa6376
2. P. Ostroumov, C. Contreras, A. Plastun, J. Rathke, T. Schultheiss, A. Taylor et al., Elliptical superconducting rf cavities for frib energy upgrade, Nuclear Instruments and Methods in Physics Research Section A: Accelerators, Spectrometers, Detectors and Associated Equipment 888 (2018) 53. https://doi.org/10.1016/j.nima.2018.01.001
3. H. Zheng, P. Sha, J. Zhai, W. Pan, Z. Li, Z. Mi et al., Development and vertical tests of 650 mhz 2-cell superconducting cavities with higher order mode couplers, Nuclear Instruments and Methods in Physics Research Section A: Accelerators, Spectrometers, Detectors and Associated Equipment 995 (2021) 165093.https://doi.org/10.1016/j.nima.2021.165093
4. S.-H. Kim, R. Afanador, D. Barnhart, M. Crofford, B. Degraff, M. Doleans et al., Overview of ten-year operation of the superconducting linear accelerator at the spallation neutron source, Nuclear Instruments and Methods in Physics Research Section A: Accelerators, Spectrometers, Detectors and Associated Equipment 852 (2017) 20.220. https://doi.org/10.1016/j.nima.2017.02.009
5. M. Martinello, D.J. Bice, C. Boffo, S.K. Chandrasekeran, G.V. Eremeev, F. Furuta et al., Q-factor optimization for high-beta 650MHz cavities for PIP-II, Journal of Applied Physics 130 (2021) 174501. https://doi.org/10.1063/5.0068531
6. D. Reschke, V. Gubarev, J. Schaffran, L. Steder, N. Walker, M. Wenskat et al., Performance in the vertical test of the 832 nine-cell 1.3 ghz cavities for the european x-ray free electron laser, Phys.

`
Rev. Accel. Beams 20 (2017) 042004.225.
https://link.aps.org/doi/10.1103/PhysRevAccelBeams.20.042004
7. T. Behnke, J.E. Brau, B. Foster, J. Fuster and M. Harrison, The International Linear Collider Technical Design Report -Volume 1, Executive Summary, 1306.6327
8. A. Grassellino, A. Romanenko, D. Sergatskov, O. Melnychuk, Y. Trenikhina, A. Crawford et al., Nitrogen and argon doping of niobium for superconducting radio frequency cavities: a pathway to highly efficient accelerating structures, Superconductor Science and Technology 26 (2013) 102001. https://dx.doi.org/10.1088/0953-2048/26/10/102001
9. S. Posen, A. Romanenko, A. Grassellino, O. Melnychuk and D. Sergatskov, Ultralow surface resistance via vacuum heat treatment of superconducting radio-frequency cavities, Phys. Rev. Appl. 13 (2020) 014024. https://link.aps.org/doi/10.1103/PhysRevApplied.13.014024
10. H. Ito, H. Araki, K. Takahashi and K. Umemori, Influence of furnace baking on Q−E behavior of superconducting accelerating cavities, Progress of Theoretical and Experimental Physics 2021(2021) 071G01.236. https://doi.org/10.1093/ptep/ptab056
11. F. He, W. Pan, P. Sha, J. Zhai, Z. Mi, X. Dai et al., Medium-temperature furnace baking of 1.3 Ghz 9-cell superconducting cavities at ihep, Superconductor Science and Technology 34 (2021) 095005. https://dx.doi.org/10.1088/1361-6668/ac1657
12. P. Sha, W.-M. Pan, S. Jin, J.-Y. Zhai, Z.-H. Mi, B.-Q. Liu et al., Ultrahigh accelerating gradient and quality factor of cepc 650mhz superconducting radio-frequency cavity, Nuclear Science and Techniques 33 (2022) 125. https://doi.org/10.1007/s41365-022-01109-8
13. T. Powers, Theory and Practice of Cavity RF Test Systems, in Proceedings of the 12th International Workshop on RF Superconductivit, 2005, https://accelconf.web.cern.ch/SRF2005/papers/sup02.pdf.
14. F. Liwen, W. Fang, L. Lin and J. Hao, Development of digital self-excited loop in vertical tests of superconducting cavity, High Power Laser and Particle Beams 33 (2021) 024001. http://www.hplpb.com.cn/article/doi/10.11884/HPLPB202133.200216
15. W. Chang, Y. He, L. Wen, C. Li, Z. Xue, Y. Song et al., A vertical test system for china-ads project injector ii superconducting cavities, Chinese Physics C 38 (2014) 057001. https://dx.doi.org/10.1088/1674-1137/38/5/057001
16. O. Melnychuk, A. Grassellino and A. Romanenko, Error analysis for intrinsic quality factor measurement in superconducting radio frequency resonators, Review of Scientific Instruments 85(2014) 124705.250. https://doi.org/10.1063/1.4903868
17. T. Powers and M. Morrone, Jefferson Lab Vertical Test Area RF System Improvement, in Proceedings of SRF2015, 2015, https://accelconf.web.cern.ch/SRF2015/papers/tupb094.pdf.
18. W. Schappert, J.P. Holzbauer, Y.M. Pischalnikov and D.A. Sergatskov, Systematic Uncertainties in RF-Based Measurement of Superconducting Cavity Quality Factors, in Proceedings of SRF2015,2015, https://accelconf.web.cern.ch/SRF2015/papers/tupb091.pdf.
19. J. Holzbauer, C. Contreras, Y. Pischalnikov, D. Sergatskov and W. Schappert, Improved rf measurements of srf cavity quality factors, Nuclear Instruments and Methods in Physics Research Section A: Accelerators, Spectrometers, Detectors and Associated Equipment 913 (2019) 7. https://doi.org/10.1016/j.nima.2018.09.155

20. V.A. Goryashko, A.K. Bhattacharyya, H. Li, D. Dancila and R. Ruber, A method for high-precision characterization of the ☐-slope of superconducting rf cavities, IEEE Transactions on Microwave Theory and Techniques 64 (2016) 3764. https://doi.org/10.1109/TMTT.2016.2605671
21. D. Frolov, Intrinsic quality factor extraction of multi-port cavity with arbitrary coupling, Review of Scientific Instruments 92 (2021) 014704. https://doi.org/10.1063/5.0014471
22. M. Jin-Ying, Q. Feng, S. Long-Bo, Z. Zheng-Long, J. Tian-Cai, X. Zong-Heng et al., Precise calibration of cavity forward and reflected signals using low-level radio-frequency system, Nuclear Science and Techniques 33 4.   https://doi.org/10.1007/s41365-022-00985-4
23. M. Jin-Ying, X. Cheng-Ye, W. An-Dong, J. Guo-Dong, T. Yue, X. Zong-Heng et al., Measurement of the cavity-loaded quality factor in superconducting radio-frequency systems with mismatched source impedance, Nuclear Science and Techniques 34 123. https://doi.org/10.1007/s41365-023-01281-5
24. J. Delayen, Phase and amplitude stabilization of superconducting resonators, Ph.D. thesis, Caltech,1978

TABLES

**Table 1.** Values of input variables.

| variable | $C_{in}$ | $C_r$ | $C_t$ | $P_{xm}$ | $\tau$ |
|---|---|---|---|---|---|
| Error Input(%) | 6.64 | 5.14 | 6.4 | 4.2 | 4 |

FIGURE CAPTIONS

**Fig. 1.** Schematic of VTS at IHEP.

**Fig. 2.** Control cabinets of VTS at IHEP.

**Fig. 3.** Vertical test results of 650 MHz SRF cavity.

**Fig. 4.** Fractional uncertainties on $E_{acc}$, $Q_0$ and $\beta^*$ as a function of $E_{acc}$.

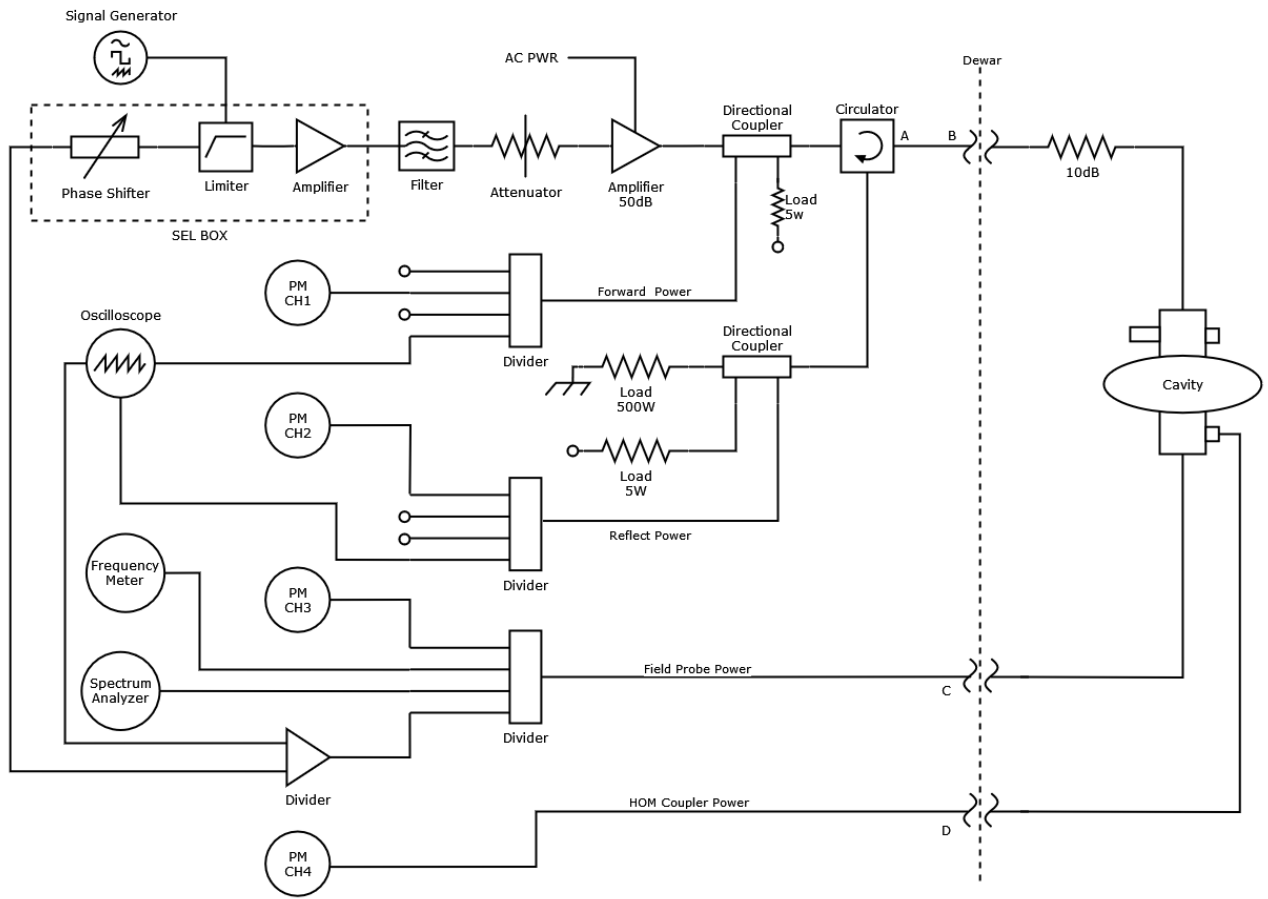

Fig. 1.

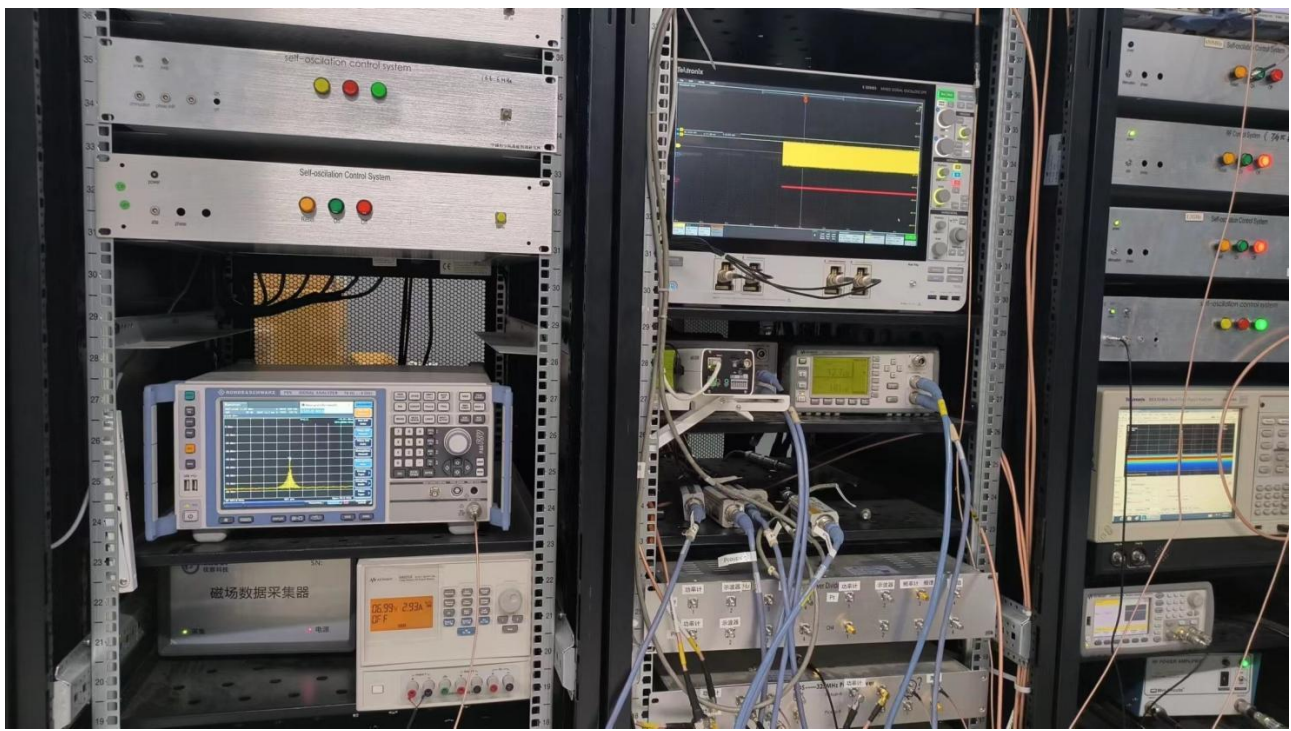

Fig. 2.

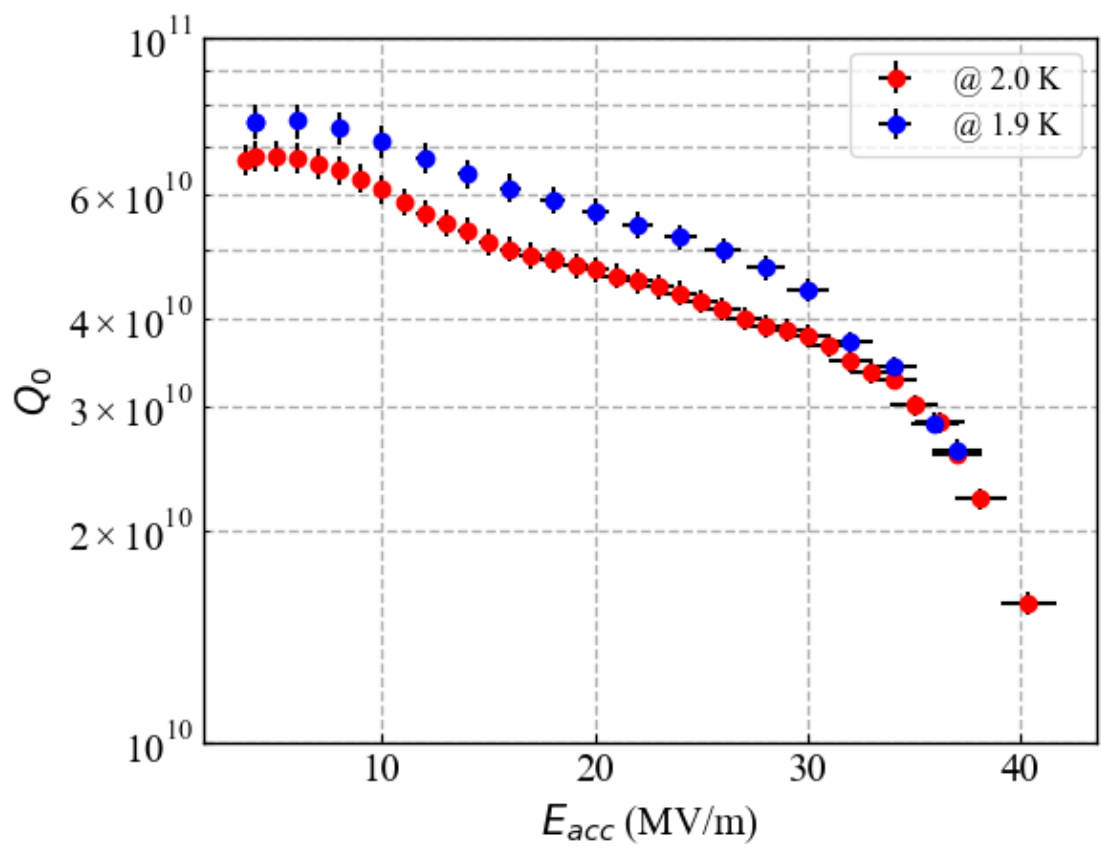

Fig. 3.

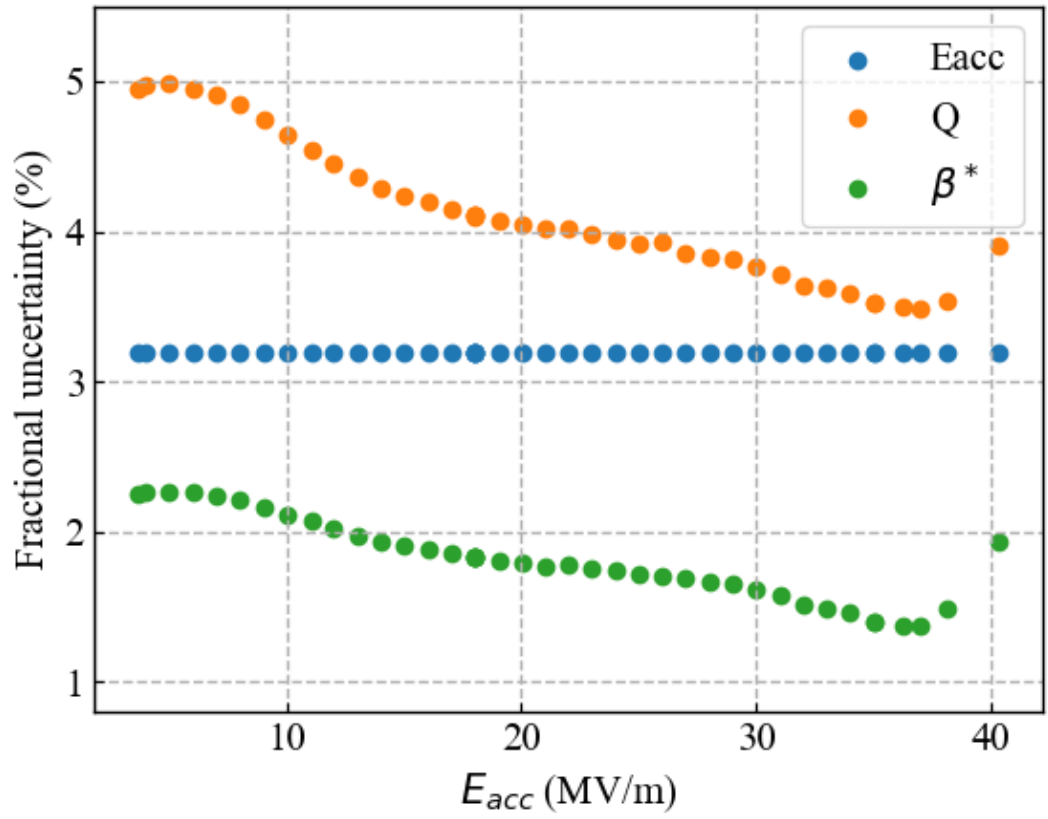

Fig. 4.